\documentstyle[preprint,aps,graphicx]{revtex} 
\begin{document}
\draft
\preprint{}
\title{ $R_b $ in Supergravity Grand Unification \\
with Non-universal Soft SUSY Breaking}
\author{ Tarakeshwar Dasgupta and Pran  Nath}
\address{Department of Physics, Northeastern University\\
Boston, MA  02115}
\date{\today}
\maketitle

\begin{abstract}
An analysis of supersymmetric contributions to $R_b$ in supergravity
grand unification with non-universal 
boundary conditions on soft SUSY breaking in the scalar 
sector is given.  Effects on $R_b$ of 
Planck scale corrections  on gaugino masses are also analysed. It is 
found that there exist regions of the parameter space where positive 
corrections to  $R_b$ of size $\sim 1 \sigma$  
can be gotten. The region of the parameter space where enhancement 
of $R_b$ occurs is identified. Prediction of the full sparticle 
spectrum for the maximal $R_b$ case is given. The analysis has  
implications for the discovery of supersymmetric particles at colliders.
\end{abstract}
\noindent
The Standard Model(SM) predicts a value of $R_b$ of 
\begin{equation}
R_b^{SM}=0.2159, m_t=175 \, {\rm GeV}
\end{equation}
with $\delta R_b/\delta m_t=-0.0002$. The experimental value of $R_b$ has
been in a state of flux over the past couple of years. 
The experimental analyses
in 1994-95  indicated $R_b = 0.2208\pm 0.0024$. 
However, more recently $R_b^{exp}$ has  drifted
downwards, and currently, assuming the value of $R_c$ at its  SM value of 
$R_c=0.172$, one finds \cite{lep} 
\begin{equation}
R_b^{exp}=0.2178\pm 0.0011
\end{equation}
which is about $1.8\sigma$ higher than the SM value.
The possibility of a discrepancy between $R_b^{exp}$ and the 
 SM value has  aroused much interest, since if valid
 the result would  signal the onset of new physics beyond the SM.
There have been several analyses recently to understand the possible origin
of potentially large $R_b$ corrections. Specifically 
supersymmetric contributions to this process have been analysed 
 within MSSM \cite{boulw,wells,garcia,chank,wang}.
A variety of other suggestions have also been made, such as corrections
from additional $Z'$ and from additional fermion generations.

In this Letter we give the first analysis of the maximal
 SUSY corrections  within supergravity unification\cite{can,applied} 
with radiative breaking of the electroweak symmetry with
 non-universal boundary conditions\cite{soni,matallio,nonuni}
 including 
Planck scale corrections to the gauge kinetic energy function in
supergravity \cite{das}. 
For comparison with the previous work we also give results for
the  maximum SUSY
corrections in MSSM, and in minimal supergravity. 
One defines 
$R_b=\Gamma(Z\rightarrow b\bar b)/\Gamma(Z\rightarrow {\rm hadrons})$
and the supersymmetric corrections to $R_b$ by 
$R_b=R_b^{SM}(m_t,m_b)$+$\Delta R_b^{SUSY}$,
where $\Delta R_b^{SUSY}$ can be written in the form\cite{boulw}
\begin{eqnarray}
\Delta R_b^{SUSY}&&=R_b^{SM}(0,0)[1-R_b^{SM}(0,0)] \nonumber \\
&&\times [\nabla_b^{SUSY}(m_t,m_b)
-\nabla_b^{SUSY}(0,0)]
\end{eqnarray}
Numerically $R_b^{SM}(0,0)$=0.2196  and $\nabla_b^{SUSY}(m_t,m_b)$
is given by 
\begin{equation}
\nabla_b^{SUSY}(m_t,m_b)=\frac{\alpha}{2\pi \sin^2\theta_W}
(\frac{v_L F_L+v_R F_R}{(v_L)^2+(v_R)^2})
\end{equation}
where $v_L$ is defined by $v_L=-\frac{1}{2}+\frac{1}{3}\sin^2\theta_W$, and
$v_R$ by $v_R=\frac{1}{3}\sin^2\theta_W$.
In supersymmetric models the quantities F$_{L,R}$ receive
 one loop contributions from 
the charged Higgs, the chargino, the neutralinos and  the gluino. The most 
dominant terms are those arising from the chargino
exchange and we exhibit these below\cite{boulw}

\begin{eqnarray}
&&F_{L,R}^{\tilde W} =(B_1^{\alpha i}v_{L,R}
-\frac{4}{3} \sin^2\theta_W C_{24}^{i\alpha i})\Lambda_{i\alpha}^{L,R}
\Lambda_{i\alpha}^{*L,R} \nonumber \\
&&+C_{24}^{i\alpha j} T_{i1}^*T_{j1}
\Lambda_{i\alpha}^{L,R}
\Lambda_{j\alpha}^{*L,R}+
M_{\tilde W_{\alpha}}M_{\tilde W_{\beta}}
C_0^{\alpha i\beta}O_{\alpha \beta}^{L,R} \Lambda_{i\alpha }^{L,R}
\Lambda_{i\beta}^{*L,R} \nonumber \\
&&+(2C_{24}^{\alpha i\beta}-M_Z^2(C_{12}^{\alpha i\beta} C_{23}^{\alpha i\beta})
-\frac{1}{2})
O_{\alpha \beta}^{R.L} \Lambda_{i\alpha }^{L,R}
\Lambda_{i\beta}^{*L,R}
\end{eqnarray}
where $\alpha, \beta (i,j)$ are the chargino(stop) indices.
$B_1,C_0,C_{12}$ etc  are given in terms of the  Passarino-Veltman
functions\cite{passa} 
and  $\Lambda_{i\alpha}^{L,R}$ are given by 
\begin{eqnarray}
\Lambda_{i\alpha}^L&=&T_{i1} V_{\alpha 1}^*-\frac{m_t}{\sqrt{2} M_W \sin\beta} 
T_{i2}V_{\alpha 2}^*\\
\Lambda_{i\alpha}^R&=&-\frac{m_b}{\sqrt{2} M_W \cos\beta} T_{i1}U_{\alpha 2}^*
\end{eqnarray}
where tan$\beta=\langle H_2\rangle /\langle H_1\rangle$ is the ratio of the Higgs VEV's, 
 $O_{\alpha\beta}^{L,R}$ are defined by
$O_{\alpha\beta}^L$=$-\cos^2\theta_W$$ \delta_{\alpha\beta}$+$\frac{1}{2}$
 $U^*_{\alpha 2}U_{\beta 2}$
and
$O_{\alpha\beta}^R$=$-\cos^2\theta_W$ $\delta_{\alpha\beta}$+$\frac{1}{2}$ 
$V^*_{\alpha 2}V_{\beta 2}$,
 where U$_{\alpha\beta}$, V$_{\alpha\beta}$ are the matrices  that diagonalize the 
chargino mass matrix and T$_{ij}$ is the matrix that diagonalizes the 
stop mass$^2$ matrix, i.e, 

\begin{equation}
\left( {\tilde t_2  }\atop{\tilde t_1}\right)=
\left(
{{ {\cos\theta_{\tilde t} }\atop{ -\sin\theta_{\tilde t}}}
{{ \sin\theta_{\tilde t}} \atop { \cos\theta_{\tilde t}}   }}
\right)
\left( {\tilde t_L  }\atop{\tilde t_R}\right)
\end{equation}

To set the stage for the analysis in supergravity unification we discuss first 
the general features  that lead to a large $\Delta R_b$ contribution in SUSY
models.
The maximum contribution to $\Delta R_b^{SUSY}$ comes from the terms involving light
masses in Eq. (5).  So a large $\Delta R_b^{SUSY}$ will require light 
$\tilde \chi_1^{\pm}, \tilde t_1$ \cite{boulw,wells,garcia,chank,wang} and 
 for low tan$\beta$ $M_{\tilde \chi_{2}^{\pm}} \approx 
M_{\tilde \chi_{1}^{\pm}}$\cite{chank} which is possible for 
$M_2 \approx -\mu$, where 
$M_2$ is the SU(2) gaugino mass and $\mu$ is the Higgs mixing parameter(
for an overview of large tan$\beta$ case  
see Ref.\cite{chank}).
Further,  $\Lambda^{L}_{ij}$'s and $O^{L,R}_{ij}$'s that give large
weights to the dominant terms in Eq.(5) lead to a large $\Delta R_b^{SUSY}$.
Large weights for the dominant terms 
require a large $\Lambda_{11}^L (\Lambda_{12}^L)$ and a large
negative $O_{11}^{L,R} (O_{22}^{L,R})$ implying a large 
$T_{12}, V_{12}(V_{22})$ and a small $U_{12}(U_{22})$ for 
$\tan\beta<1(\tan\beta>1)$.  We find that for $\tan\beta>1$  
$\Delta R_b^{SUSY}$
is maximum for $\theta_{\tilde t} \simeq -9^o$  and a   $\tilde \chi_{2}^{\pm}$ 
which is mixture of a large up-higgsino
and a gaugino state($|V_{22}|>0.9,|U_{22}| <0.1$).
Our results are in accord with the analysis of Ref.$\!\!$\cite{chank} except 
for $\theta_{\tilde t}^{opt.}$ where our value supports the result of
Ref.$\!\!$\cite{boulw}.
Our best value of 
$\Delta R_b$ in MSSM then is $\Delta R_b^{SUSY}\leq 0.0028$ for
$\tan\beta\geq 1.16$, comparable with previous 
determinations\cite{wells,garcia,chank,wang}.

	Although, as discussed above, one can generate a 
	significant $\Delta R_b^{SUSY}$ correction in MSSM, it is not 
	a priori clear what part of the parameter space, if any, which gives
	large corrections is compatible with the constraints of 
	grand unification and radiative breaking of the electro-weak symmetry.  
	This is the issue we address in this Letter.
The analysis we carry out includes radiative breaking of the electroweak
symmetry, constraints to avoid color and charge breaking,  
 experimental constraints on the superparticle spectrum
 and the $b\rightarrow s+\gamma $ experimental constraint as given by
 the CLEO Collaboration\cite{cleo}. 
We also include the constraint arising from the decay 
$t\rightarrow $$\tilde t_1 \tilde\chi_i^0$
and  assume that the branching ratio of the top decay into stops  satisfies
B(t$\rightarrow $$\tilde t_1 \tilde\chi_i^0)<0.4$.
We discuss first the minimal supergravity case which is 
 parameterized by  $m_0$,$m_{1/2}$, $A_0$ and tan$\beta$, 
where  $m_0$ is the universal scalar mass, $m_{1/2}$ is the universal gaugino
mass, and $A_0$ is the universal trilinear coupling.
We find that the maximal supersymmetric contribution to R$_b$ 
is $\Delta R_b^{SUSY}=0.0002$ over the entire parameter space investigated.
Our  result is in accord with previous analyses\cite{wang}
 where it was also found that
the minimal supergravity grand 
unification does not produce a significant correction to $R_b$. 
\\

The rest of this Letter is devoted to a discussion of R$_b$ in supergravity
unification with non-universal soft SUSY breaking. 
While the simplest supergravity models are based on universal soft SUSY
breaking, the general framework of the theory\cite{can,applied}
 allows for the existence
of non-universalities via a generational dependent Kahler potential\cite{soni}.  
The non-universalities that affect $R_b$ most sensitively are the 
 non-universalities in the Higgs sector and in the third generation 
sector. For this reason we shall focus in the present analysis on 
the non-universalities in these sectors 
and assume universality in the remaining sectors. It has recently been shown
that the non-universalities in the Higgs  sector and in the third generation
sector are strongly coupled because  of the
large top Yukawa coupling \cite{nonuni}. This phenomenon will 
play an important role in our analysis.
It is convenient to parameterize the 
non-universalities in the Higgs sector by $\delta_{H_1}, \delta_{H_2}$ where
$m_{H_1}^2(0)=m_0^2(1+\delta_{H_1})$, and $m_{H_2}^2(0)=m_0^2(1+\delta_{H_2})$.
Similarly we parameterize the non-universalities in the third generation 
sector by $\delta_{\tilde t_L}$ and $\delta_{\tilde t_R}$ where
$m_{\tilde t_L}^2(0)=m_0^2(1+\delta_{\tilde t_L})$, and 
$m_{\tilde t_R}^2(0)=m_0^2(1+\delta_{\tilde t_R})$.
We also include in the analysis Planck scale corrections which arise via
corrections to the gauge kinetic energy, i.e, 
 $ -\frac{1}{4}\it f_{\alpha\beta}F_{\mu\nu}^{\alpha}F^{\beta \mu\nu}$,
where $\it f_{\alpha \beta}$ contains the 
corrections from the Planck scale. Planck corrections in 
$\it f_{\alpha \beta}$ contribute to gauge
coupling unification in supergravity GUT\cite{das} and also generate 
corrections to the gaugino masses which can be parameterized by
$M_i$=$\frac{\alpha_i}{\alpha_G}$$(1+c'\frac{M}{M_P}n_i)$$m_{\frac{1}{2}}$,
where M is the GUT mass, $M_P$ is the Planck mass, $c'$ parameterizes the Planck 
scale correction and $n_i$ are subgroup indices\cite{das}.
Thus for the non-minimal model we have
the set of parameters 
$c'$, $\delta_{H_1}$, $\delta_{H_2}$, $\delta_{\tilde t_L}$, and  
$\delta_{\tilde t_R}$,
in addition to the parameters of the minimal model.

In Fig.(1) we display $R_b$ in the Standard Model and the maximal
$R_b$  that can be achieved in supergravity models  
with universal and non-universal boundary conditions. As discussed above 
the supersymmetric contributions
for the universal case are always small, maximally 
$\Delta R_b^{SUSY}=0.0002$. 
However, for the non-universal case one can get much larger contributions.
Thus for $\mu<0$ the maximal $\Delta R_b^{SUSY}$ is $0.0011$,
and for  $\mu>0$ the maximal $\Delta R_b^{SUSY}$ is 0.0008. 
As discussed earlier the maximal  $\Delta R_b^{SUSY}$ is associated with a 
relatively light 
chargino ${\tilde{\chi^{\pm}}_1}$ and  a relatively light stop 
${\tilde t_1}$. In Fig.(2) we display the correlation between the
light chargino mass  and the light stop mass 
for the maximal $\Delta R_b^{SUSY}$ for the case $\mu<0$ and a similar analysis 
holds for the case $\mu>0$. We find that the maximal 
$\Delta R_b^{SUSY}$ decreases systematically with increasing mass 
of the light stop and the light chargino, and one cannot maintain a
$\sim 1\sigma$ correction to the SM value with both the light stop
and the light chargino above 100 GeV. Thus if the experimental lower limits on 
the light chargino and the light stop exceed 100 GeV, then the maximal 
$\Delta R_b^{SUSY}$  in supergravity grand unification with inclusion 
of non-universalities is not in excess
of 0.0006. Further, if both the chargino and the light Higgs lie above 
100 GeV, then $\Delta R_b^{SUSY}$ reduces to a value similar to what one
has in the minimal case.

 We have also computed the full
supersymmetric spectrum for some typical cases where $\Delta R_b^{SUSY}$ is
large. 
 We  exhibit in Table 1 the mass spectra of the 
 supersymmetric particles which maximize  
R$_b$ for 6-discrete sets of chargino-stop masses for the $\mu<0$ case. 
We find that in all cases 
$\delta_{H_2} = -\delta_{H_1} = 1.15 - 1.17$ and 
$\delta_{\tilde{t}_L} \simeq -\delta_{\tilde{t}_R} = 0.25 - 0.35$.
The relative signs of the non-universalities, i.e., opposite 
signs for  $\delta_{H_1}$ and $\delta_{H_2}$ and for  
$\delta_{\tilde{t}_L}$ and  $\delta_{\tilde{t}_R}$ 
can be easily understood
by looking at the non-universality correction to $\mu^2$ and to the
stop masses. The correction to $\mu^2$ is given by\cite{nonuni}
$\Delta\mu^2$=$(t^2-1)^{-1}$$(\delta_{H_1}$-$(\delta_{H_2}$+$\frac{D_0-1}{2}$
$\delta)t^2)$,
where t=tan$\beta$,$\delta=\delta_{H_2}+\delta_{\tilde t_L}+
\delta_{\tilde t_R}$, and $D_0=0$ 
determines  the position of the Landau pole singularity\cite{nonuni}. 
For $m_t$=175 GeV,
$M_G=10^{16.2}$ GeV, one has $D_0=0.27$. One finds then that a $\delta_{H_1}
<0$ and a $\delta_{H_2}>0$ gives a negative contribution to $\mu^2$ 
and makes $|\mu|$ small, which is what is needed as can be seen in Table 1.
  The correction to $m_{\tilde t_L}^2$  and $m_{\tilde t_R}^2$  are given 
  by \cite{nonuni}
$\Delta m_{\tilde t_L}^2$ = $m_0^2$$(\delta_{\tilde t_L}$ + 
$\frac{D_0-1}{6}$$\delta)$, and 
$\Delta m_{\tilde t_R}^2$ =$ m_0^2$$(\delta_{\tilde t_R}$ + 
$\frac{D_0-1}{3}$$\delta)$.
Here for values of  $\delta_{H_2}$, $\delta_{\tilde t_L}$ and 
$\delta_{\tilde t_R}$ indicated, e.g. for 
 $\delta_{H_2}=1.15$, $\delta_{\tilde t_L}=-\delta_{t_R}=0.25$, 
one finds  $\Delta m_{\tilde t_L}^2$ = 0.11$m_0^2$
and $\Delta m_{\tilde t_R}^2$ = -0.53 $m_0^2$. Since $\tilde t_1\approx
\tilde t_R$, one finds then that the sign of the non-universalities is such 
 as to split the $\tilde t_1-\tilde t_2$ masses, making $\tilde t_1$ 
 lighter and $\tilde t_2$ heavier. This effect enhances the value of 
 $R_b$. 
The analysis shows that most of the corrections to $R_b$  
come from the non-universalities in the scalar sector, and $c'$ is seen not to 
play a significant role, i.e., the 
effect of $c'$ on $\Delta R_b^{SUSY}$ is less  
than $5\%$. In Fig. 3 we present the same analysis as in Fig. 1 but as a 
function of the top mass.
Numerical results for m$_t$=175 GeV are summarized in Table 2. 
We note that the upper limit of $R_b^{SUSY}\leq 0.0011$ in the non-universal 
case
is mostly due to the fact that one needs a high value of $\tan\beta$ to obtain 
low values of $M_{\tilde t_1}, \mu$ and $M_{\tilde \chi_1^{\pm}} $ 
using radiative
breaking.

Prediction of the sparticle spectrum in the non-universal supergravity 
model depends on the size of $\Delta R_b^{SUSY}$ one assumes. If one 
requires a sizable $\Delta R_b^{SUSY}$ correction, which we take  here
to imply a correction greater than $0.0006$, i.e., greater than 
$\sim \frac{1}{2}\sigma$,
 then for both signs of $\mu$ 
the light Higgs will have 
a mass below 93 GeV, the light chargino and the light stop will have masses
 around or below 100 GeV, and the gluino mass will lie below 450 GeV(525 GeV)
 for $\mu<0$($\mu>0$). Thus the entire range of the light chargino and 
 the light stop masses will be fully accessible at the Tevatron in the Main 
Injector era. Since the light  Higgs lies below 93 GeV in this case, it must be
visible at LEP II if it achieves its optimum energy of $\sqrt{s}$=192 GeV
and an integrated luminosity of $500$ pb$^{-1}$ or at  TeV33  
with 5-10 fb$^{-1}$ of integrated luminosity\cite{amidei}. Regarding the gluino essentially 
the entire gluino mass 
range for $\mu<0$ and the range up to 450 GeV for $\mu>0$
could be probed at TeV33  with an integrated 
luminosity of 100 fb$^{-1}$\cite{amidei}. Thus the supergravity model with
$\Delta R_b^{SUSY}>0.0006$  can
be completely tested in the Higgs, chargino and stop  sectors for both signs
of $\mu$ at LEP II and at TeV33. It can also be completely(partially) 
tested in the gluino sector 
for $\mu<0(\mu>0)$ at TeV33. 
If no SUSY particles are  seen in the mass ranges indicated, 
then $\Delta R_b^{SUSY}$ must lie below the level of 0.0006, i.e., below
$\sim \frac{1}{2} \sigma$.

 In this letter we have given the first analysis of the maximal $R_b$ that 
 can be gotten in supergravity unification with non-universal boundary
 conditions on the soft SUSY breaking parameters.  We find maximal
  $\Delta R_b^{SUSY}\simeq 1\sigma(0.8\sigma)$ for $\mu<0(>0)$ which is 
  significantly smaller than the maximum value one can get in MSSM but 
  significantly larger than the maximum value achievable in minimal
  supergravity unification. Thus values of $\Delta R_b^{SUSY}$ in MSSM
  in excess of 0.0011(0.0008) for $\mu<0(\mu>0)$ are in conflict with 
  the twin constraints of  grand unification and radiative breaking of the
  electro-weak symmetry. 
   The $\Delta R_b$ supergravity correction gives a correction to the
   LEP value of $\alpha_s$ of $\Delta \alpha_s= -4\Delta R_b$ 
  which amounts to a
maximal correction of  $\Delta \alpha_s=-0.0044(-0.0032)$ for 
$\mu<0(\mu>0)$. Recalling the discrepancy between the LEP value
of $\alpha_s$ and the DIS value of $\alpha_s$\cite{shifman}, 
one finds that 
supergravity unification  with non-universal soft SUSY breaking 
can bridge the gap maximally only half way between the LEP value and the
DIS value of $\alpha_s$.   
   The analysis makes several  
 predictions on the sparticle spectra which can be tested at colliders
 in the near future. The analysis on maximal $\Delta R_b^{SUSY}$ presented
 here is also applicable to the class of string models which have the SM
 gauge group  and no extra generations below the GUT scale. 
 
{\bf Acknowledgements}:
This research was supported in part by NSF grant PHY-96020274.\\

\pagebreak
\begin{center}
{\bf TABLES\\}
{\small Table 1: Mass spectra of supersymmetric particles.\\}
{\small Table 2: Maximal $\Delta R_b^{SUSY}$ in models vs experiment. 
The last four entries are from this analysis. For other 
determinations of  $\Delta R_b^{SUSY}$ in MSSM  see 
Refs. \cite{wells,garcia,wang}.\\}
\end{center}
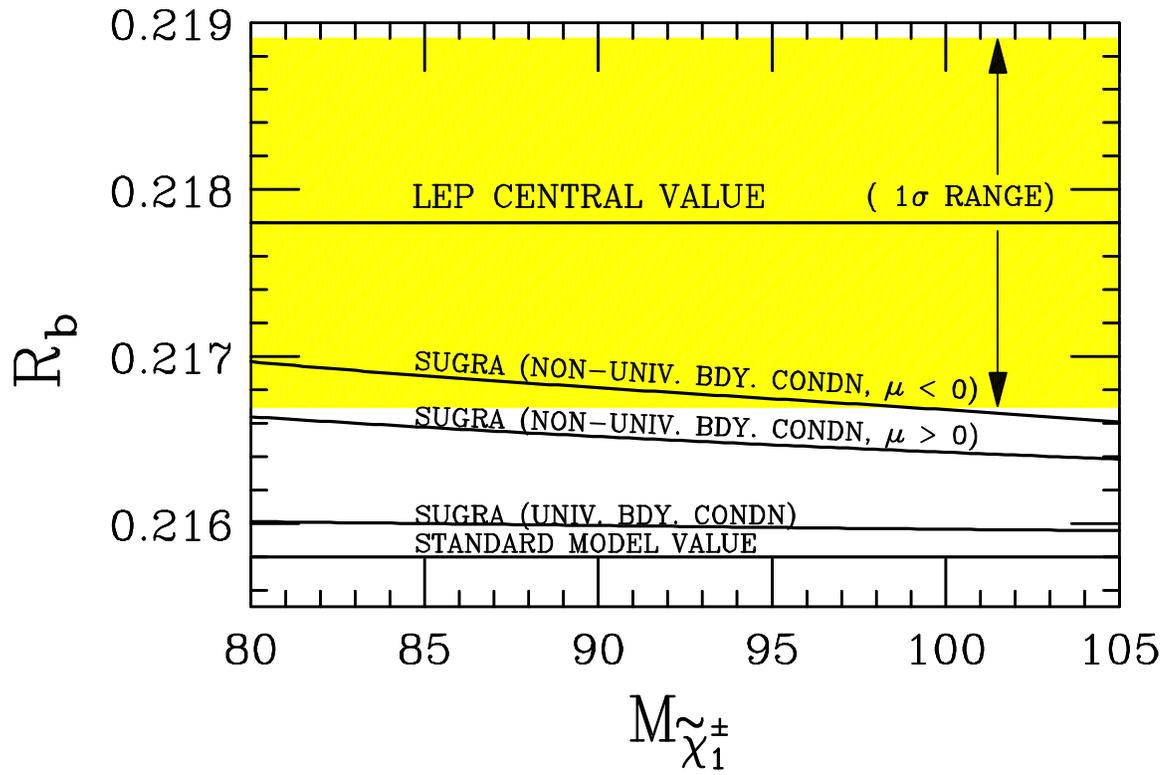
\begin{figure} 
\caption{\small Maximum $R_b$ for various models  as a function of
the light $m_{\tilde\chi_1^{\pm}}$.\\}
\end{figure}
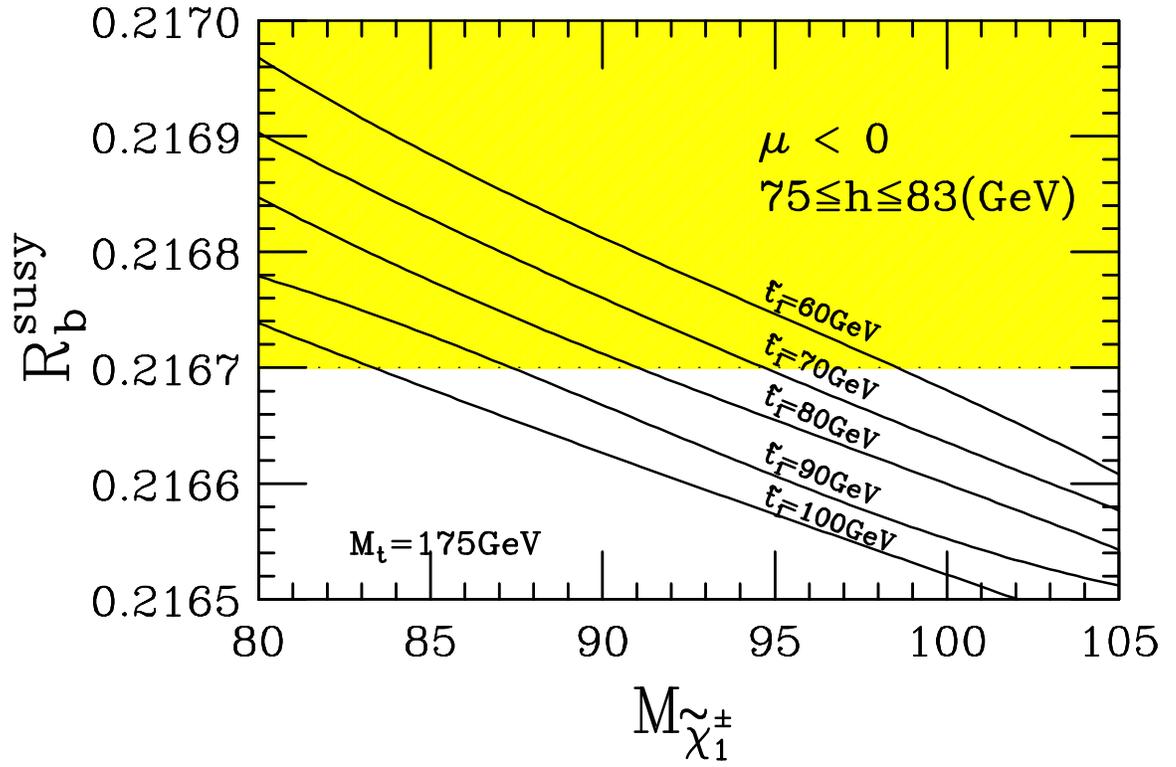
\begin{figure} 
\caption{\small Maximum $R_b$ as a function of
$m_{\tilde\chi_1^{\pm}}$ for 
different $m_{\tilde t_1}$ with  
non-universalities.\\}
\end{figure}
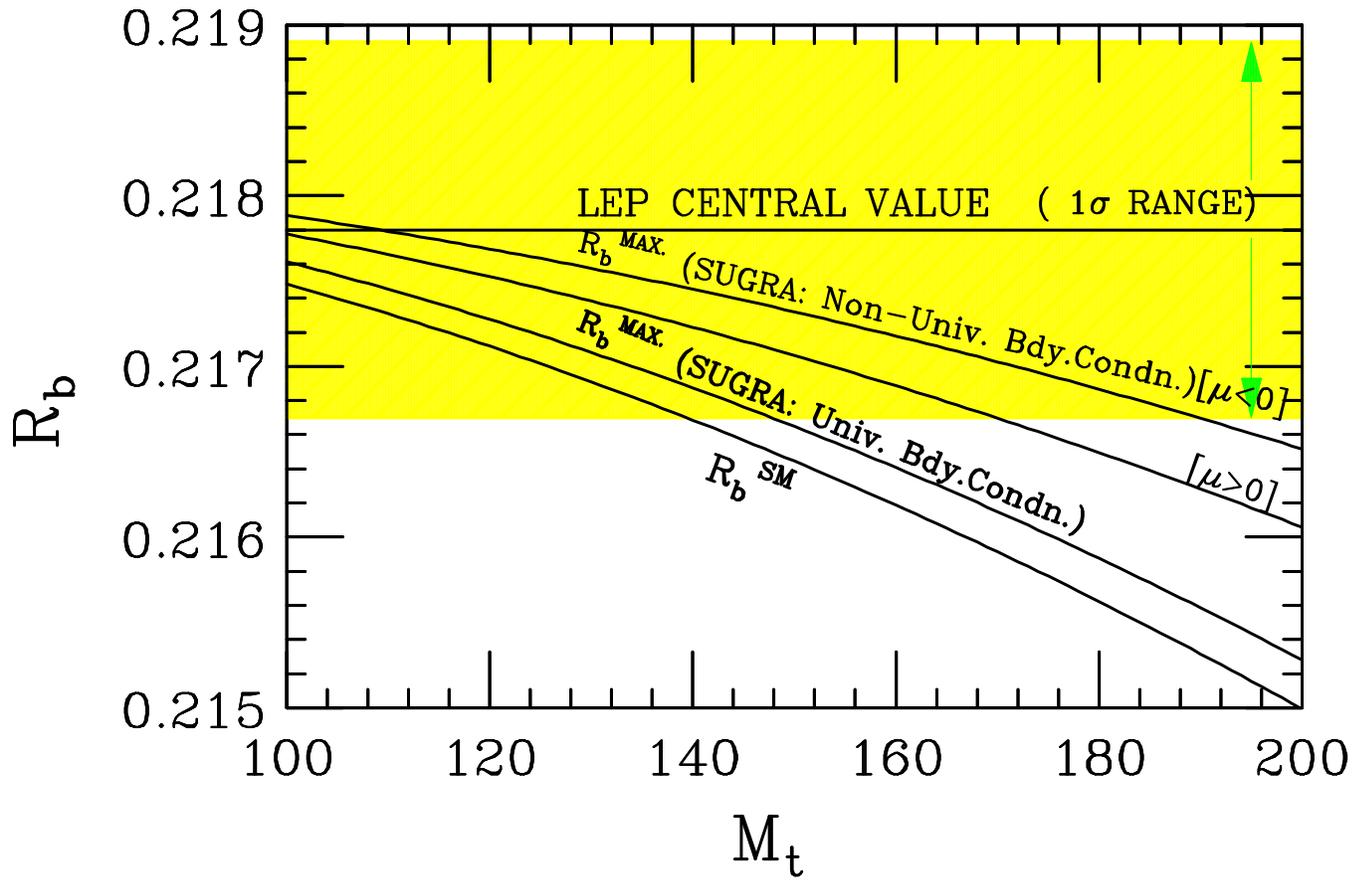
\begin{figure} 
\caption{\small Maximum $R_b$ as a function of $M_t$ for different models.\\}
\end{figure}

\end{document}